\documentclass[a4paper]{spie}  

\usepackage[]{graphicx}
\usepackage{epsfig}

\title{Dissipation-assisted quantum computation in  atom-cavity systems}
\author{Almut Beige, Hugo Cable, and Peter L. Knight
\skiplinehalf
Blackett Laboratory, Imperial College London, Prince Consort Road, London SW7 2BW, UK}
\date{\today}

\authorinfo{Further author information contact \\ H.C.: E-mail: hugo.cable@imperial.ac.uk\\  A.B.: E-mail: a.beige@imperial.ac.uk\\ P.L.K.: E-mail: p.knight@imperial.ac.uk }

\begin{document}
\maketitle

\begin{abstract}
The principal obstacle to quantum information processing with many qubits is decoherence.  One source of decoherence is spontaneous emission which causes loss of energy and information.  Inability to control system parameters with high precision is another possible source of error.  Strategies aimed at overcoming one kind of error typically increase sensitivity to others.  As a solution we propose quantum computing with {\em dissipation-assisted quantum gates}. These can be run relatively fast while achieving fidelities close to one. The success rate of each gate operation can, at least in principle, be arbitrary close to one.
\end{abstract}


\section{Introduction}

Quantum computing is a paradigm in which quantum entanglement and interference are exploited for information processing.  Quantum algorithms have been exhibited which are exponentially better than the best known classical solutions\cite{shor,deutsch,grover}.  Elements of quantum computing have been demonstrated experimentally using nuclear magnetic resonance, trapped atoms and ions, superconductivity and non-linear optical technologies\cite{exps}.  However building systems with many coupled qubits remains a huge challenge.  Many demands must be met: reliable qubit storage, preparation and measurement, gate operations with high fidelity and low failure rate, scalability, practical time scales and accurate transportation or teleportation of states.  The biggest problem is posed by decoherence which tends to destroy the desirable quantum behaviour.

\begin{figure}
\begin{center}
\begin{tabular}{c}
\includegraphics[height=2.5cm]{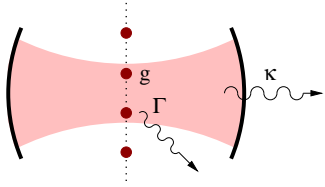}
\end{tabular}
\end{center}
\caption[example]{ \label{fig1} 
Schematic view of an atom-cavity system. Two atoms are moved into the resonator where a two qubit gate is performed by simultaneous application of laser fields. The coupling of each atom to the cavity mode is $g$ while $\Gamma$ and $\kappa$ are spontaneous decay rates from the atoms and the cavity, respectively.}
\end{figure}

In this paper we review ways to overcome the problem of decoherence in quantum information processing with many qubits. As an example we consider cold atoms trapped in an optical cavity as shown in Figure \ref{fig1}. They provide a promising technology for quantum computing as well as an ideal model for theoretical studies. Problems that must be overcome are common to other implementations. The quantum optics of atoms in cavities is well established. Each qubit is obtained from two different atomic ground states of one atom. To implement a two qubit gate, the two corresponding atoms are moved inside the optical resonator where both see the same coupling constant. The key sources of decoherence are dissipation of cavity excitation with a rate $\kappa$ and spontaneous decay from excited atomic levels with decay rate $\Gamma$.

In recent years many schemes have been proposed for entangled state preparation\cite{plenio,beige4,SorensenMolmer,marr} and two qubit gates\cite{letter,ZhengGuo,BeigeTregenna,PachosWalther,JanePlenioJonathan,Grangieretal,YiSuYou} for atoms in optical cavities. Since it is difficult to construct a system of high quality optical cavities proximity to the so called {\em bad-cavity limit}, where the atom-cavity coupling constant $g$ is about the same size as the spontaneous decay rates,
\begin{equation}
g \sim \Gamma \sim \kappa ~,
\end{equation}
provides a fundamental basis for comparison of schemes. Several other criteria may be identified for practicality and scalability.  A scheme should be simple and relatively insensitive to the precise values of the system parameters. Operations must be reliable with a fidelity close to one while the success rate of an operation can be significantly less than one as long as errors are detected and the scheme is repeated as necessary.

Several schemes try to solve the dissipation problem by avoiding the population of excited states with the help of strong detunings\cite{ZhengGuo,JanePlenioJonathan,YiSuYou} or an environment-induced quantum Zeno effect\cite{beige4,letter,BeigeTregenna}. Other schemes work only with a success rate well below $50 \, \%$\cite{plenio,SorensenMolmer}.  A problem shared by many of these schemes is that their operation time is much longer than the inverse atom-cavity coupling constant. Thus, while the probability for one type of dissipation is made very small, the probability for failure due to the other type increases significantly. For this reason, the quantum computing scheme proposed by Pellizzari {\em et al.}\cite{pellizzari} in 1995  is still one of the best as regards dissipation. Demanding a certain minimum fidelity of the prepared state, it requires the smallest ratio between $g^2$ and $\kappa \Gamma$. 

Intuition suggests that decoherence is always damaging for quantum computation. On the contrary, we show here that dissipation can be exploited to achieve coherent control of open quantum systems with high fidelity and success rate. To do so we review the idea of {\em quantum computing using dissipation}\cite{Zurek,letter,BeigeTregenna} and compare a quantum computing scheme based on this idea with a scheme based on {\em dissipation-assisted adiabatic passages}. It is shown that dissipation can stabilise the time evolution of an open quantum system. Under special circumstances, dissipation can speed up gate operations that can only be performed much slower in the absence of spontaneous decay rates.  For other schemes based on dissipation-assisted adiabatic passages see References\cite{PachosWalther,marr,Beige}. 

As concretes example we consider schemes for the realisation of a CNOT gate, a controlled PHASE and a two qubit SWAP gate which can be executed in one step by applying laser fields. The schemes are widely independent from the value of the atom-cavity coupling constant $g$. While being simpler and less sensitive to parameter fluctuations, the proposed quantum computing scheme is comparable with other schemes\cite{PachosWalther,pellizzari} with respect to the amount of dissipation that can be allowed.

\section{Non-Hermitian Hamiltonians and no-photon time evolution} \label{sect:2}

In this section a theoretical model for describing the time evolution of open quantum systems under the condition of no dissipation is introduced. The type of dissipation we consider here results from the loss of energy due to spontaneous photon emission and we give a short review of the quantum jump method\cite{HeWi11} which is equivalent to the Monte Carlo wave-function\cite{HeWi2} and the quantum trajectory\cite{HeWi3} approaches \cite{review}.

\subsection{The quantum jump approach}

Stretching back three decades, several quantum optical experiments have been performed studying the statistics of photons emitted by one or two laser-driven trapped atoms or ions and effects have been found that would be averaged out in the statistics of photons emitted by a whole ensemble. One example is electron shelving\cite{shelving}, the occurance of macroscopic light and dark periods in the resonance fluorescence of a laser-driven atom with a metastable state. Another is the two-atom double-slit experiment by Eichmann {\em et al.}\cite{Eichmann} which demonstrated that the photons emitted by two atoms at a fixed distance can create an interference pattern on a distant screen. These experiments suggest that the effect of the environment on the state of the atoms is the same as the effect of rapidly repeated measurements of whether a photon has been emitted or not\cite{behe,schoen}. From this assumption the quantum jump approach has been derived\cite{HeWi11}. 

Assume that a measurement is performed on a quantum optical system surrounded by a free radiation field initially in its vaccum state $|0_{\rm ph} \rangle$ and prepared in $|\psi \rangle$ determining after a time $\Delta t$ whether or not a photon has been created.  If $H$ is the Hamiltonian including the interaction of the system with its environment, the state of the system equals
\begin{equation} \label{def}
|0_{\rm ph} \rangle \, U_{\rm cond}(\Delta t,0) |\psi \rangle
\equiv    |0_{\rm ph} \rangle  \langle 0_{\rm ph} | \, U(\Delta t,0) \, |0_{\rm ph} \rangle |\psi \rangle 
\end{equation}
under the condition that the free radiation field is still in the vacuum state. For quantum optical experiments, it has been shown that the dynamics under the conditional time evolution operator $U_{\rm cond}(\Delta t,0)$, defined by the right hand side of Equation (\ref{def}), can be summarised by a Hamiltonian $H_{\rm cond}$ that is largely independent of the choice of $\Delta t$. The conditional Hamiltonian $H_{\rm cond}$ is non-Hermitian and the norm of a state vector developing with $H_{\rm cond}$ decreases in general in time such that
\begin{equation} \label{P0}
P_0(t,\psi) = \| \, U_{\rm cond} (t,0) \, |\psi \rangle \, \|^2 
\end{equation}
equals the probability for no photon emission in $(0,t)$. 

\subsection{Dissipation-assisted quantum computation}

The time evolution of  a quantum system under the condition of no photon emission is coherent. It can be described using the Schr\"odinger equation with the conditional Hamiltonian $H_{\rm cond}$. {\em Quantum computing using dissipation}\cite{letter,BeigeTregenna} or {\em dissipation-assisted quantum computation}\cite{Beige,PachosWalther} aims to exploit  the conditional time evolution to implement better gate operations. Whenever a photon emission occurs the computation fails and the experiment has to be repeated. Nevertheless, quantum gates can be implemented with fidelities $F$ very close to unity and success rates $P_0$ well above $90 \, \%$.

The time evolution of a system with decay rates can differ significantly from the time evolution of a system with no decay rates, even if no actual photon emission takes place. This can be explained by the fact that observation of no photons leads to a continuous gain of information about the state of the system. The longer no photons are emitted the more unlikely it is that the system has population in excited states that might cause an emission. In the formalism this is taken into account by the non-Hermitian terms in the conditional Hamiltonian. They continuously damp away amplitudes of unstable states\cite{Cook,Hegerfeldt}. 

For dissipation-assisted quantum computation to work and to minimise the probability of gate failure, only decoherence-free states\cite{Palma,Zanardi97,Lidar98,Guo98} should be populated. These are states whose population does not lead to a photon emission. A state $|\psi \rangle$ is decoherence-free if
\begin{equation} \label{p0cond}
P_0(t,\psi) \equiv 1
\end{equation}
for all times $t$\cite{beige4}. Hence, the decoherence-free subspace of a system is spanned by the eigenvectors of the conditional Hamiltonian $H_{\rm cond}$ with real eigenvalues. 

Sections \ref{sect:4} and \ref{sect:5} give concrete examples of mechanisms which restrict the time evolution of two atoms inside an optical resonator to a subspace decoherence-free with respect to leakage of photons through the cavity mirrors. In general it is not possible to find a subspace that is decoherence-free with respect to all error sources at the same time. The best that can be done is to suppress one type of dissipation and to operate the scheme in a parameter regime minimising the total error rate. 
The key strength of quantum gates based on the no-photon time evolution of a system is that the fidelity of the finally prepared state can be very close to one. This is the case if unwanted states are non decoherence-free and the non-Hermitian terms in the conditional Hamiltonian continuously decrease their amplitudes. 

The trade off for high fidelities is usually an increase of the probability for failure of the proposed scheme due to dissipation. However, the non-Hermitian terms in $H_{\rm cond}$ also inhibit transitions to unwanted states. In this way, they stabilise the desired time evolution of the system. For a wide range of parameters, this probability for no photon emission and gate success rate $P_0(t,\psi)$  can be shown to be well above $50 \,\%$. 

Let us now estimate the probability of finding the result of a whole computation assuming that each gate can be performed with maximum fidelity but only with a finite success rate. The probability of implementing an algorithm of $N$ gates faultlessly is $P_0^N$ and grows exponentially with $N$. On the other hand, if one always knows whether or not an algorithm has failed, the computation can be repeated until a result is obtained. The probability of not having a result after $M$ runs equals
\begin{equation}
P_{\rm no \, result}= \big(\, 1 - P_0^N \, \big)^M ~.
\end{equation}
For large $N$ this is approximately ${\rm e}^{-c}$ where $M=c P_0^{-N}$ and many repetitions might be necessary to implement a computation. However, for smaller $N$ and $P_0$ sufficiently close to one, the failure probability is already nearly negligible for $M \approx N$. For example, if $P_0=95\,\%$ and an algorithm with $N=50$ gates is performed, then repeating the computation 50 times yields a success rate above $98\,\%$.

\subsection{Example: Two three-level atoms inside an optical cavity} \label{sect:2.3}

\begin{figure}
\begin{center}
\begin{tabular}{c}
\includegraphics[height=2.5cm]{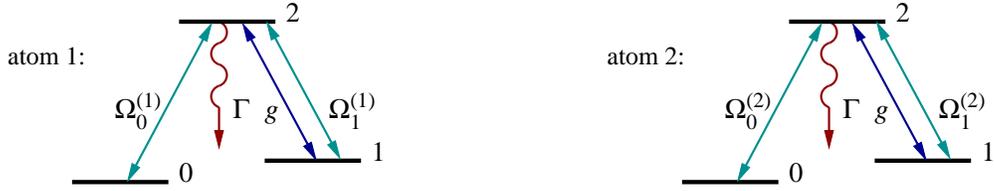}
\end{tabular}
\end{center}
\caption[example]{ \label{fig2} 
Level scheme of the two atoms inside the resonator. Each qubit is obtained from a pair of ground states $|0 \rangle$ and $|1 \rangle$ of one atom. To implement gate operations, laser fields with Rabi frequencies $\Omega_j^{(i)}$ can be applied exciting the $j$-2 transition of atom $i$.}
\end{figure}

As an example for {\em quantum computing using dissipation} and {\em dissipation-assisted quantum computation} this paper discusses two implementations of universal two qubit quantum gates with atoms in optical cavities (see Sections \ref{sect:4.2}, \ref{sect:5.2} and \ref{sect:5.3}). The two atoms involved in the gate operation should be trapped inside an optical cavity, as shown in Figure \ref{fig1}, which couples resonantly to the 1-2 transition of each atom. This can be achieved with the help of a linear ion trap as in the experiment by Guth\"orlein {\em et al.}\cite{guthoerlein} or by using a cavity mounted on an atomic chip as described in Reference\cite{horak}. Each qubit is obtained from the ground states $|0 \rangle$ and $|1 \rangle$ of the same atom and laser pulses are applied simultaneously to manipulate the state of the atoms. The atomic level configurations and laser driven transitions are shown in Figure \ref{fig2}. 

The Hamiltonian of this system in the Schr\"{o}dinger picture has components
\begin{equation}
H = H_{\rm atom} + H_{\rm laser} + H_{\rm cavity}
+H_{\rm atom-cavity}+H_{\rm cavity-env}+H_{\rm atom-env}~.
\end{equation}
Proceeding as in Reference\cite{plenio} and assuming that the environment continuously monitors whether photons leak out of the cavity or from excited atomic states, the conditional Hamiltonian for the two atoms inside the cavity can be derived.  Let us denote the Rabi frequency of the laser interacting with the $j$-2 transition of atom $i$ as $\Omega_j^{(i)}$ while $b$ and $b^\dagger$ are the annihilation and creation operators for a single photon inside the cavity mode. Then the conditional Hamiltonian in the interaction picture with respect to the interaction free Hamiltonian $H_{\rm atom}+H_{\rm cavity}$  is
\begin{equation} \label{cond}
H _{\rm cond} = {\rm i} \hbar g \sum_{i=1}^2 b^\dagger |1\rangle _j \langle2| + 
\sum_{i=1}^2 \sum_{j=0}^1 {\textstyle {1 \over 2}} \hbar \Omega^{(i)}_j  |j \rangle_{ii} \langle 2 | + {\rm h.c.} 
-  {\textstyle {{\rm i} \over 2}} \hbar \kappa b ^{\dagger} b 
-  {\textstyle {{\rm i} \over 2}} \hbar \Gamma \sum_i |2 \rangle_{ii} \langle 2| ~.
\end{equation}
The last two contributions to this equation are the non-Hermitian terms. 

From Equation (\ref{cond}) one can now easily calculate the decoherence-free states of the atoms inside the optical cavity with respect to leakage of photons through the cavity mirrors. They include all superpositions of the atomic ground states, i.e. the qubits states of the system and also the maximally entangled atomic state 
\begin{equation}
|a \rangle \equiv {\textstyle {1 \over \sqrt{2}}} \, (|12 \rangle - |21 \rangle) 
\end{equation}
while the cavity is in its vacuum state $|0 \rangle_{\rm cav}$. Prepared in the state $|0 \rangle_{\rm cav} |a \rangle  \equiv |0;a\rangle$ the atoms do not interact with the cavity mode ($H_{\rm atom-cavity}|a;0\rangle =0$)  and therefore cannot transfer their excitation into the resonator. Thus no photon can leak out through the cavity mirrrors. One can easily prove that the decoherence-free states of the system are indeed the eigenstates of the conditional Hamiltonian with real eigenvalues and for $\Gamma=0$.

\section{Quantum computing using dissipation} \label{sect:4}

In this section we look in detail at how a quantum Zeno effect arising from interaction with the environment can be exploited for quantum computing.  Theoretical schemes for two qubit gates with dissipation and spontaneous decay can achieve several desirable properties simultaneously: near perfect fidelity, a success rate close to one and insensitivity to the precise values of system parameters which are hard to control in experiments.  As an example we describe a scheme for a CNOT gate\cite{letter,BeigeTregenna} implemented using two atoms trapped in an optical cavity.  A particular strength of this scheme is that it can be implemented in one step.  
  
\subsection{The basic idea} \label{sect:4.1}

Suppose that all non decoherence-free states of a system couple strongly to the environment and that populating them typically leads to a photon emission within $\Delta t$. The time $\Delta t$ is greater than a certain minimal size which can be determined from the quantum jump approach with the help of the conditional time evolution operator $U_{\rm cond}(\Delta t,0)$. Note that the eigenvectors of $H_{\rm cond}$ are in general non-orthogonal. Using the reciprocal basis $\{ |\lambda^k \rangle \}$ with  $\langle \lambda^k |\lambda_j \rangle = \delta_{jk}$, the operator $H_{\rm cond}$ can be written as 
\begin{equation}
H_{\rm cond} = \sum_k \lambda_k \, |\lambda_k \rangle \langle \lambda^k| ~.
\end{equation}
Provided that eigenvalues $\lambda_k$ of $H_{\rm cond}$ satisfy 
\begin{equation} \label{ass}
{\rm e}^{-{\rm i} \lambda_k \Delta t/\hbar} = 0 ~~
{\rm for~all} ~ k ~ {\rm with} ~ {\rm Im} \, \lambda_k \neq 0 ~,
\end{equation}
the conditional time evolution operator becomes
\begin{equation} \label{ups}
U_{\rm cond}(\Delta t,0) = \sum_{i: |\lambda_i \rangle \in {\rm DFS}}
{\rm e}^{-{\rm i} \lambda_i \Delta t/\hbar} \, |\lambda_i\rangle \langle \lambda_i| ~.
\end{equation}
This operator projects onto the decoherence-free subspace and one can interpret the action of the free radiation field over a time $\Delta t$ as a measurement whether the system is decoherence-free or not. Within $\Delta t$, the population of purely non decoherence-free states causes an emission with certainty. The probability for no photon emission in $\Delta t$ is exactly the probability to be in a decoherence-free state. 

A weak interaction acting over a time scale much longer than $\Delta t$ can realise a coherent time evolution inside the decoherence-free subspace.  This can be interpreted as a quantum Zeno effect\cite{behe,misra}.  In $\Delta t$ the weak interaction can create at worst a population proportional to $\Delta t$ outside the decoherence-free subspace where it would be registered with probability $\Delta t^2$ over the next interval.  Therefore the probability of always finding the population in the decoherence-free subspace, $T/\Delta t$ times over the gate operation time $T$, tends to one in the limit of weak measurement.  All transitions out of the decoherence-free subspace are strongly inhibited.        

The time evolution within the decoherence-free subspace is not inhibited and can be summarised with the help of the effective Hamiltonian\cite{beige4}
\begin{equation} \label{heff}
H_{\rm eff} = I\!\!P_{\rm DFS} \, H_{\rm cond} \, I\!\!P_{\rm DFS} ~,
\end{equation}
where $H_{\rm cond}$ includes amongst its terms an interaction realising a gate operation and 
\begin{equation}
I\!\!P_{\rm DFS} = \sum_{i:|\lambda_i\rangle \in {\rm DFS}}
|\lambda_i\rangle \langle \lambda_i|
\end{equation}
is the projection onto the decoherence-free subspace.

The basic idea of {\em quantum computing using dissipation} is to utilise the effective time evolution (\ref{heff}) for the implementation of gate operations.  The decoherence-free space typically contains in addition to all the ground states, states which are highly entangled.  By populating these entanglement between qubits can be created even if the interaction Hamiltonian is not an entangling one.  In this way it is much easier to make quantum information processors in the laboratory.   

Deviations from the effective time evolution occur if the applied interaction is not weak enough and a small population exists outside of the decoherence-free subspace during the no-photon time evolution. During such an evolution a photon emission might take place. We denote the corrections to the desired time evolution by $-U_{\rm corr}(T,0)$. Then,  
\begin{equation}
U_{\rm cond} (T,0) = U_{\rm eff}(T,0) - U_{\rm corr}(T,0) ~.
\end{equation}
From (\ref{P0}) the gate success rate $P_0(T,\psi)$ equals
\begin{equation} \label{uhu}
P_0(T,\psi) = 1 - 2 \,
{\rm Re} \, \langle \psi | U_{\rm eff}(T,0) U_{\rm corr}(T,0) |\psi \rangle ~.
\end{equation}
Initially prepared in a decoherence-free state, the fidelity of gate operations under the condition of no photon emission is given by
\begin{equation} \label{F}
F(T,\psi) = {| \langle \psi | U_{\rm eff}(T,0) U_{\rm cond} (T,0) |\psi \rangle |^2 \over P_0(T,\psi)} ~.
\end{equation}
Using Eq.~(\ref{uhu}), it can be shown that the fidelity $F$ equals one in first order $U_{\rm corr}(T,0)$ even if the gate success rate $P_0(T,\psi)$ differs from one. In conclusion, the quantum Zeno effect can be used to implement quantum gates with a high precision.

\subsection{Example: CNOT gate}  \label{sect:4.2}

If the time evolution of the system is as predicted in the previous subsection then the population is restricted to the decoherence-free subspace and the time evolution of the atoms is given to a very good approximation by the effective Hamiltonian (\ref{heff}). For the setup described in Section \ref{sect:2.3} this equals 
\begin{equation} \label{heff2}
H_{\rm eff} = {\textstyle {1 \over 2 \sqrt{2}}} \, \hbar \big[ \, 
- \Omega_0^{(1)} \,  |01 \rangle \langle a| + \Omega_0^{(2)} \, |10 \rangle \langle a|  
+ \big( \Omega_1^{(2)} - \Omega_1^{(1)} \big) \,  |11 \rangle \langle a| \, \big]  + {\rm h.c.} 
\end{equation}
and is achieved by working in the regime $\Omega_j^{(i)} \ll  g^2/\kappa$ and $\kappa$\cite{letter,BeigeTregenna}. A number of two qubit quantum gates can be realised with this Hamiltonian.  By populating $|a\rangle$ one can entangle the two atoms.  Hence gates such as the universal two qubit CNOT and PHASE are possible. The gate operation time $T$ needs to be chosen such that at the end of each operation again only qubit states are populated. 

\begin{figure}
\begin{center}
\begin{tabular}{c}
\includegraphics[height=4.5cm]{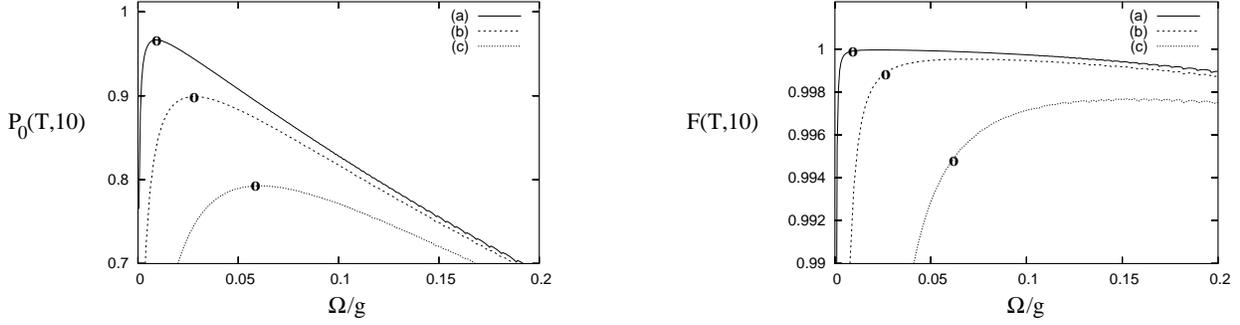}
\end{tabular}
\end{center}
\caption[example]{ \label{fig6} 
Success rate $P_0(T,\psi)$ and fidelity $F(T,\psi)$ for a CNOT gate with $\kappa=1$; $\Gamma=0.0001 \, g$ (a), $\Gamma=0.001 \, g$ (b) and $\Gamma=0.005 \, g$ (c); and initial qubit state $|10 \rangle$. For all other qubit states, the success rate and fidelity are higher and are actually one for the initial qubit state $|01 \rangle$. The circles mark positions where the no-photon probability is maximal and the corresponding worst fidelity. A short transition time at the end of each gate has been assumed during which the amplitudes of states with the damping rates $\kappa$ and $\Gamma$ disappear.}
\end{figure} 

As an example we now discuss the realisation of a CNOT gate with the help of a single laser pulse. If atom 1 contains the target bit and atom 2 provides the control bit, then the CNOT gate is given by the unitary
\begin{equation}\label{ucnot}
U_{\rm CNOT} = |00\rangle \langle00| + |01\rangle \langle01| + |10\rangle \langle11| + |11\rangle \langle10| ~.
\end{equation}
This is most easily achieved with one laser field coupling to the 1-2 transition of atom 1 with Rabi frequency $\Omega \equiv \Omega_1^{(1)}$. Another laser field should couple with the same Rabi frequency to the 0-2 transition of atom 2. This corresponds to the effective Hamiltonian
\begin{equation}
H_{\rm eff} = {\textstyle {1 \over 2 \sqrt{2}}} \, \hbar \Omega \, \big[ \, |10 \rangle \langle a|
- |11 \rangle \langle a| \, \big]  + {\rm h.c.} 
\end{equation}
The required laser pulse duration is $T = 2\pi/\Omega$ with a small correction to allow the population remaining on level $|0a \rangle$ at the end of the gate to damp away. 

The numerical results presented in Figure \ref{fig6} bear out the theory. Since gates working in the quantum Zeno effect regime are intrinisically slow they are vulnerable to spontaneous decay.  Hence the constraint $ \Gamma \ll \Omega$ must also be satisfied and $\Gamma$ has to be below $0.005 \, g$. On the other hand they allow the cavity decay rate $\kappa$ to be of about the same size as the coupling constant $g$. Optimising $\Omega$, the no-photon probability can be as high as $79 \, \%$ if $g^2= 200 \, \kappa \Gamma$, as high as $89 \, \%$ if $g^2 = 1 \, 000 \, \kappa \Gamma$ and above $96 \, \%$ for $g^2 > 10 \, 000 \, \kappa \Gamma$. The main disadvantage of quantum computing schemes based on an environment-induced quantum Zeno effect is that they are very slow. This makes them more susceptible to errors such as dephasing.

\section{Dissipation-assisted quantum gates} \label{sect:5}

The main advantage of the quantum computing scheme discussed in Section \ref{sect:4} is that 
a great variety of gates can be realised by choosing the Rabi frequencies $\Omega_j^{(i)}$ appropriately. In this section we show that this flexibility can be exchanged for a decrease of the gate operation time by about two orders of magnitude. In the following fixed laser configurations are chosen and a simple scheme for the implementation of the two qubit PHASE and SWAP gates are presented. 

\subsection{The basic idea}
  
Dissipation plays the key role in quantum computing schemes based on an environment-induced quantum Zeno effect. As explained in Sections 2 and 3, it continuously inhibits the transitions into non decoherence-free states by damping away the amplitudes of states that can decohere. However, we have found that for some particular Rabi frequencies, schemes that work in the Quantum Zeno effect regime also work in the absence of dissipation $(\kappa=0)$.  

These schemes work for completely different reasons to those described in Section \ref{sect:4}. For small Rabi frequencies $\Omega_j^{(i)}$ the key role is now played by adiabaticity arising from two time scales. These different time scales are provided by the atom-cavity constant $g$ being a few orders of magnitude larger than the Rabi frequencies $\Omega_j^{(i)}$ of the applied laser fields. In the absence of spontaneous decay rates, amplitudes of states undergoing a fast time evolution adapt immediately to the amplitudes of the slowly varying states and follow their time evolution. This can be shown by solving the time evolution with the help of an adiabatic elimination (see subsection \ref{sect:5.2}) and results in the same effective Hamiltonian as given by Equation (\ref{heff2}).  
 
In the quantum Zeno parameter regime, these gates work with the same fidelities $F$ and success rates $P_0$. This is because gates based on both effects can be analysed using an adiabatic elimination of fast varying coefficients. While it is not possible to operate schemes as the CNOT gate outside the adiabatic regime, these others can be performed much faster with Rabi frequencies $\Omega_j^{(i)}$ of about the same size as $g$. The prepared state has a very high overlap with the state predicted by the effective time evolution (\ref{heff2}) and fidelities above $99 \, \%$ are achieved easily. For a wide range of parameters, the gate success rates are well above $50 \, \%$.

The role of dissipation in these schemes is to stabilise the adiabatic time evolution by damping away population that accumulates due to non-adiabaticity in unwanted states. Dissipation acts like an error detection measurement, especially when a short transition time is allowed for the amplitudes of non-decoherence-free states to damp away. Since the time evolution of the system is as expected for an adiabatic process, it can be called a {\em dissipation-assited adiabatic passage}. For other examples of dissipation-assisted quantum gates with cold trapped ions or for atoms in optical cavities see References\cite{PachosWalther,Beige}.

\subsection{Example: PHASE gate}  \label{sect:5.2}

An example for a universal two qubit gate based on a dissipation-assited adiabatic passage is the PHASE gate which changes the state $|1\rangle$ of the second atom into $-|1\rangle$ provided that the first atom is in $|0\rangle$. This gate can be implemented with a single laser pulse by choosing $\Omega_0^{(1)} \equiv \Omega$ and $\Omega_0^{(2)} =\Omega_1^{(1)} =\Omega_1^{(2)}  = 0$. In contrast to the previous section, let us now assume that the spontaneous decay rates are zero $(\kappa = \Gamma =0)$. The Hamiltonian of the laser-driven atoms inside the cavity equals 
\begin{equation} \label{nocond}
H  = {\rm i} \hbar g \sum_{i=1}^2 b^\dagger |1\rangle _j \langle2| + 
{\textstyle {1 \over 2}} \hbar \Omega  |0 \rangle_{11} \langle 2 | + {\rm h.c.} 
\end{equation}
in the interaction picture. In the following we derive the effective time evolution of the system under the condition that $\Omega \ll g$.

\begin{figure}
\begin{center}
\begin{tabular}{c}
\includegraphics[height=3.5cm]{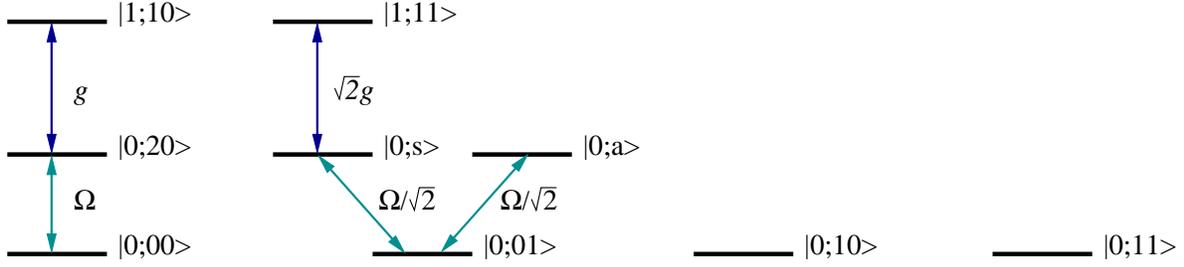}
\end{tabular}
\end{center}
\caption[example]{ \label{fig5} 
Level configuration showing all transitions involved in the implementation of a PHASE gate with a single laser pulse.}
\end{figure} 

We denote by  $|n;x \rangle$ a state with $n$ cavity photons and the atoms in $|x \rangle$. The state $|x \rangle$ can be a state $|ij \rangle$ $(i,j=0,1,2)$ or one of the maximally entangled states $|a \rangle$ and 
\begin{equation}
|s \rangle \equiv {\textstyle {1 \over \sqrt{2}}} \, (|12 \rangle + |21 \rangle) ~. 
\end{equation}
Starting from a qubit state, the only states that are involved during a gate operation are $|0;00\rangle$, $|0;01\rangle$, $|0;10\rangle$, $|0;11\rangle$, $|0;20\rangle$, $|0;a\rangle$, $|0;s\rangle$, $|1;10\rangle$ and $|1;11\rangle$ (see Figure \ref{fig5}). The slowly varying amplitudes are the coefficients of the qubit states and auxiliary state $|0;a\rangle$. Proceeding as in Reference\cite{letter} and setting the derivatives of all other coefficients equal to zero gives a solution for the effective time evolution of the system including first order corrections in $\Omega/g$. To a good approximation the amplitudes $c_{n;x}$ of the fast varying states
equal
\begin{equation} \label{xxxx}
c_{0;20} =  c_{0;s} =0 ~, ~   c_{1;10} = - {\rm i}\Omega/(2g) \, c_{0;00} ~~ {\rm and} ~~    
c_{1;11} = - {\rm i}\Omega/(4g) \, c_{0;01}
\end{equation}
while the time evolution of the decoherence-free and slowly-varying states is given by the differential equations 
\begin{equation} \label{dgl}
\dot c_{0;10} =  \dot c_{0;11} =0 ~, ~   
\dot c_{0;01} = {\textstyle {1 \over  2\sqrt{2}}} \, {\rm i} \Omega  c_{0;a} ~~ {\rm and} ~~    
\dot c_{0;a} =  {\textstyle {1 \over  2\sqrt{2}}} \, {\rm i} \Omega  c_{0;01} ~.
\end{equation}
A comparison with equation (\ref{heff2}) shows that the time evolution of the system can effectively be  summarised by the Hamiltonian $H_{\rm eff}$, as expected.  The effective Hamiltonian (\ref{heff2}) becomes in this case
\begin{eqnarray} \label{hopt}
H_{\rm eff} &=& - {\textstyle {1 \over 2 \sqrt{2}}} \, \hbar \Omega \, |01 \rangle \langle a| + {\rm h.c.}
\end{eqnarray}
and the operation time $T$  should equal $2 \sqrt{2} \pi/\Omega$. The laser pulse then transfers excitation in $|01 \rangle$ into the bus state $|a \rangle$ and back --- a process during which the state vector accumulates a minus sign.  

\begin{figure}
\begin{center}
\begin{tabular}{c}
\includegraphics[height=4.5cm]{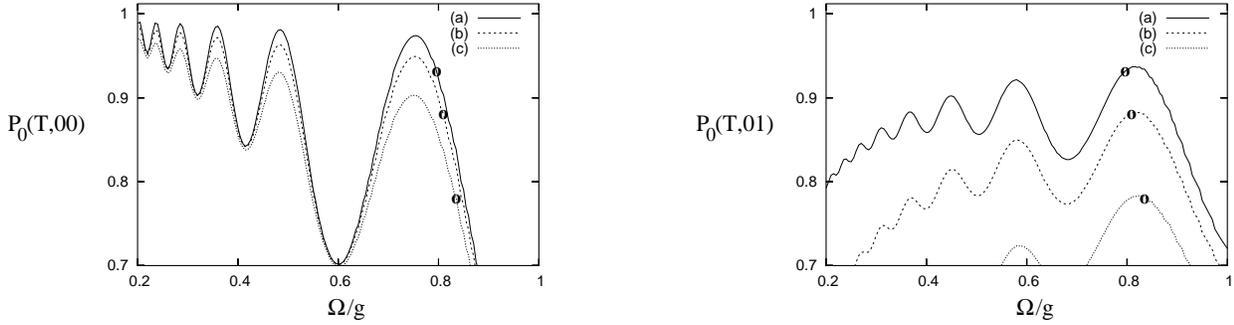}
\end{tabular}
\end{center}
\caption[example]{ \label{fig4} 
Success rate $P_0(T,\psi)$ of a single PHASE gate for the parameters $\kappa=\Gamma=0.01 \, g$ (a), $\kappa=\Gamma=0.02 \, g$ (b) and $\kappa=\Gamma=0.04 \, g$ (c) and for the initial qubit states $|00\rangle$ and $|01 \rangle$.  The circles mark positions where $P_0$ is the same for both initial states and coincides with the minimum no-photon emission probability of the scheme for the given decay rates. For the qubit states $|10\rangle$ and $|11 \rangle$ the no-photon probability equals exactly one.
Allowing for a transition time where states with the decay rates $\kappa$ and $\Gamma$ can decay, the fidelity of the finally obtained state does not differ from unity.}
\end{figure}

The condition for adiabaticity of the time evolution of the system is $\Omega \ll g$ and the maximum gate operation time that can be achieved using adiabatic passages are about the same as for gates based on the quantum Zeno effect. One advantage of the PHASE gate is that it can be operated in a regime where $\Omega$ is about the same size as $g$ where a small cavity leakage rate is taken into account. Figure \ref{fig4} shows the results from a numerical integration of the conditional time evolution (\ref{cond}). At the end of each gate operation, a short transition time allows the amplitudes of non decoherence-free states to damp away. For the parameters in Figure \ref{fig4}, the fidelity $F$ does not differ from one. Optimising $\Omega$,  the no-photon probability can be as high as $78 \, \%$ if $g^2 = 625 \, \kappa \Gamma$, while $P_0> 88 \, \%$ requires $g^2 > 2500 \, \kappa \Gamma$ and for $P_0> 93 \, \%$ one should have $g^2 > 10000 \, \kappa \Gamma$. The combined amount of dissipation allowed in the system is about the same as in subsection \ref{sect:4.2}. 

The presence of the spontaneous decay rate $\kappa$ does not affect the adiabatic time evolution and the system behaves as predicted in (\ref{xxxx}) and (\ref{dgl}) as long as the Rabi frequency $\Omega$ is small. As the Rabi frequency increases, the condition for adiabaticity no longer applies and population accumulates in the states $|0;20 \rangle$, $|0;s \rangle$, $|1;11\rangle$ and $|1;10\rangle$.  A non-zero spontaneous decay works together with the atom-cavity coupling to damp away these amplitudes ensuring that the operation fidelity remains close to one.  However, the cost of speeding up the gate operation is a lowered success rate.  If the spontaneous decay rate is too large the success probability reduces drastically and the scheme is no longer useful.

For the realisation of the CNOT gate dissipation is crucial since it restricts the time evolution of the system to the decoherence-free subspace. A level configuration, similar to the one shown in Figure \ref{fig5}, reveals that there are many more states involved in the no-photon time evolution of the system, including states with more than one cavity excitation. If the spontaneous decay rate $\kappa$ equals zero, population is transferred into these states with a rate proportional to $\Omega$, even for small Rabi frequencies, and the fidelity of the gate operation differs significantly from one. In an adiabatic elimination on the basis of $g \gg \Omega$ this is reflected by the fact that the decoherence-free states are no longer the only slowly varying states of the system.

\subsection{Example: SWAP gate}  \label{sect:5.3}

Another quantum gate that can be implemented by dissipation-assisted adiabatic passage is the SWAP gate. It transforms the state $|01\rangle$ into $|10\rangle$ and vice versa but leaves the qubit states $|00\rangle$ and $|11\rangle$ unchanged. To implement this gate one should choose $\Omega_0^{(1)} = \Omega_0^{(2)} \equiv \Omega$ and $\Omega_1^{(1)} =\Omega_1^{(2)}  = 0$ and individual laser addressing of the atoms is not required. The effective Hamiltonian (\ref{heff2}) becomes in this case
\begin{eqnarray} \label{hopt2{hopt}}
H_{\rm eff} &=& {\textstyle {1 \over 2 \sqrt{2}}} \, \hbar \Omega \, \big[ \, - |01 \rangle \langle a| + |10 \rangle \langle a| \, \big]  + {\rm h.c.}
\end{eqnarray}
and $T$  should equal $2 \pi/\Omega$. Unlike the CNOT and PHASE gate, the SWAP operation is not universal. Nevertheless, this gate can be very useful since it exchanges the states of two qubits without that the corresponding atoms have to swap places physically. 

\begin{figure}
\begin{center}
\begin{tabular}{c}
\includegraphics[height=4.5cm]{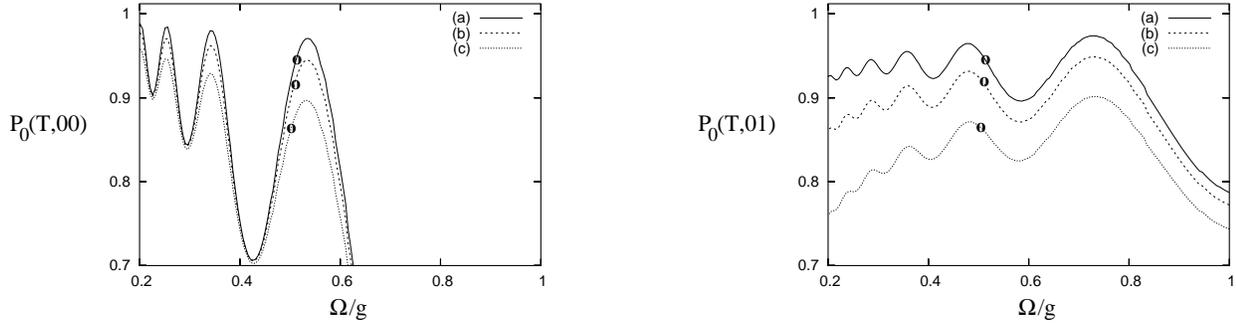}
\end{tabular}
\end{center}
\caption[example]{ \label{fig7} 
Success rate $P_0(T,\psi)$ of a single SWAP gate for the same parameters as in Figure \ref{fig4}.  For the qubit state $|11\rangle$ the no-photon probability equals exactly one while $P_0(T,01) \equiv P_0(T,10)$. Allowing for a transition time where states with the decay rates $\kappa$ and $\Gamma$ can decay, the fidelity of the finally obtained for the positions marked by circles is above $99 \, \%$.}
\end{figure} 

Figure \ref{fig7} presents the results of a numerical integration of the time evolution of a SWAP gate and shows that the success rate and fidelity are comparable to those in subsection \ref{sect:5.2}. A look at the level scheme for the SWAP operation reveals that the levels and transitions involved in the time evolution are similar to those of the PHASE gate in Figure \ref{fig5}. Spontaneous emission from the atoms affects the no-photon probability less than for the PHASE gate since the gate operation time $T$ is a factor $\sqrt{2}$ shorter. On the other hand, more population is transferred into the cavity mode since the lasers are applied to the 0-2 transition of both atoms. 

\section{Conclusions}

This paper discusses how dissipation can be used to help implement gates with fidelities close to one in quantum computing.  The focus is on atoms trapped in an optical cavity but the conclusions can be viewed more generally.  Section 2 details the governing coherent no-photon conditional time evolution described by a non-Hermitian Hamiltonian and derived from the quantum jump approach.  A scheme from which many gates may be derived with two $\Lambda$ configuration atoms and four lasers is given as an example.  

Section 3 details how dissipation in the form of a quantum Zeno effect can be used to realise quantum information processing with high fidelities and success rates and opens the door to a realm of new possibilities for implementing different kinds of quantum gate operations. The quantum Zeno effect is understood as rapidly repeated measurements of whether the population is in decoherence-free states or not. This restricts the time evolution to a decoherence-free subspace as long as laser fields are applied which are sufficiently weak. An advantage of {\em quantum computing using dissipation} is that the schemes are relatively simple to realise and the effective time evolution of the system does not depend on the exact value of the hard to control atom-cavity coupling constants as long as they are the same for both atoms. 

While quantum Zeno gates are inherently slow, we found that for specific choices of Rabi frequency, dissipation is not crucial, and the corresponding gates work as well in the absence of spontaneous decay rates. In the quantum Zeno parameter regime they work with about the same fidelities and success rates. However, these gates function in this regime because of adiabaticity arising from the different time scales set by the laser Rabi frequencies and atom-cavity coupling. They can then be operated outside the adiabatic regime with significantly shorter operation times along with less sensitivity to dephasing and other error sources. Since dissipation ensures fidelities close to one for a wide range of Rabi frequencies while the time evolution remains as predicted by adiabaticity, the process can be called a {\em dissipation-assisted adiabatic passage}. 

The concrete examples considered in the paper -- the two qubit PHASE, SWAP and CNOT gates -- have many advantages.  They are relatively simple and can be implemented in one step by applying laser fields. The demands on the cavity quality are modest. For example $g^2 > 200 \, \kappa\Gamma$ is sufficient to assure success rates around $80 \, \%$ while for $g^2 > 1 \, 000 \, \kappa\Gamma$ the success rate can be as high as $90 \, \%$.  
  
\acknowledgments
A.B. thanks the Royal Society for funding as a University Research Fellow. In addition, we would like to thank C. Marr and J. Pachos for interesting discussions. This work was supported in part by the EPSRC and the European Union.

\bibliography{report}
\bibliographystyle{spiebib}

\end{document}